\documentclass{aa}
\usepackage{natbib,graphicx}

\newcommand{\be}{\begin{equation}}
\newcommand{\ee}{\end{equation}}

\begin{document}

\date{\today}

\title{Modeling the magnetic field in the protostellar source NGC~1333~IRAS~4A}

\date{Received / Accepted}


\author{Jos\'e Gon\c{c}alves\inst{1}, Daniele Galli\inst{2}, \and Josep Miquel Girart\inst{3}}

\institute{
Centro de Astronomia e Astrof\'{\i}sica da Universidade de Lisboa, Tapada da Ajuda, 1349-018 
Lisboa, Portugal \\ 
\email{goncalve@oal.ul.pt}
\and INAF-Osservatorio Astrofisico di Arcetri, Largo E. Fermi 5, I-50125 Firenze, Italy \\
\email{galli@arcetri.astro.it}
\and {Institut de Ci\`encies de l'Espai (CSIC--IEEC), Campus
UAB--Facultat de Ci\`encies, Torre C5--Parell 2$^\mathrm{a}$, 08193 Bellaterra,
Catalunya,  Spain } \\ 
\email{girart@ieec.cat}
}

\authorrunning{Gon\c{c}alves et al.}
\titlerunning{The magnetic field in NGC~1333~IRAS~4A}

\abstract
{Magnetic fields are believed to play a crucial role in the process of star formation.}
{We compare high-angular resolution observations of the submillimeter
polarized emission of NGC~1333~IRAS~4A, tracing the magnetic field around
a low-mass protostar, with models of the collapse of magnetized molecular
cloud cores.}
{Assuming a uniform dust alignment efficiency, we computed the Stokes
parameters and synthetic polarization maps from the model density and
magnetic field distribution by integrations along the line-of-sight and 
convolution with the interferometric response.}
{The synthetic maps are in good agreement with the data. The
best-fitting models were obtained for a protostellar mass of 0.8~$M_\odot$,
of age $9\times 10^4$~yr, formed in a cloud with an initial
mass-to-flux ratio $\sim 2$ times the critical value.}
{The magnetic field morphology in NGC~1333~IRAS~4A is consistent with
the standard theoretical scenario for the formation of solar-type
stars, where well-ordered, large-scale, rather than turbulent, magnetic
fields control the evolution and collapse of the molecular cloud cores
from which stars form.}

\maketitle

\section{Introduction}
\label{intro}

\begin{figure}[t]
\resizebox{\hsize}{!}{\includegraphics[angle=-90]{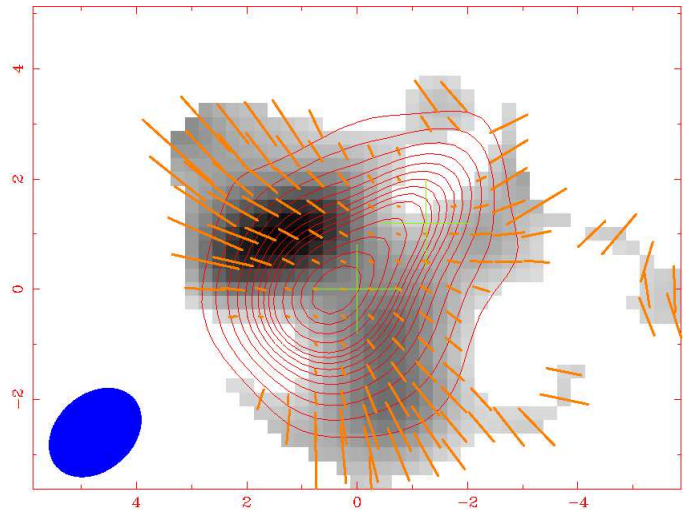}}
\caption{Map of NGC~1333~IRAS~4A, from GRM06.
Contours show the continuum emission at 877~$\mu$m, bars indicate
the direction and degree of polarization (magnetic field vectors),
and the color map shows the polarized intensity. At the distance of 
300~pc, $1\arcsec$ corresponds to 300~AU.}
\label{iras4a} 
\end{figure}

NGC~1333~IRAS~4A (herafter IRAS~4A) is one of the prototypical low-mass
young stellar systems in the earliest stages of evolution, and it is
still deeply embedded in an infalling dense molecular and dusty
envelope (Sandell et al.~1991; Di Francesco et al.~2001) and powering a
well-collimated outflow (Blake et al.~1995; Choi~2005; Choi et
al.~2006). The BIMA spectropolarimetric observations have detected and
partially resolved the polarization in both the dust (at 1.3mm) and
line (CO $J=2$--1) emission (Girart et al.~1999). Recent observations
with the submillimeter Array (SMA) at 877~$\mu$m (see
Fig.~\ref{iras4a}) show that the magnetic field associated with the
infalling envelope has a well-defined hourglass morphology on scales of
a few hundred AU (Girart et al.~2006, hereafter GRM06). In this paper,
we perform a quantitive comparison of the observed submillimeter
polarization data with models of the collapse of magnetized molecular
cloud cores and show that the data support the theoretical scenario
where the ordered, mean component of the interstellar magnetic field
controls the evolution and collapse of the molecular cloud cores from
which stars form (see e.g. Shu et al.~1987, 1999; Machida et
al.~2005a,b; Banerjee \& Pudritz~2006).

\section{The models}
\label{models}

We have adopted two models for the magnetic field in of IRAS~4A. The first is
an axisymmetric, time-dependent, calculation of the inside-out collapse
of an uniformly magnetized cloud with ambipolar diffusion (Galli \&
Shu~1993a,b, hereafter GS93). The second is a steady-state,
axisymmetric calculation of the accretion flow on a low-mass
protostar and the associated magnetic field, including the effects of
Ohmic dissipation (Shu et al.~2006, hereafter S06). Both studies
ignore the effects of rotation on the collapse (for this, see Galli et
al.~2006).

The study of GS93 focused on the formation of a large flattened
structure (``pseudo-disk'') around an accreting protostar due to the effect
of the strong non-radial component of the Lorentz force on the
collapsing matter. According to GS93, ambipolar diffusion during the
dynamical collapse phase plays a minor role on scales larger than a few
$10^2$~AU, especially at very early times after the onset of collapse.
Under quasi-field freezing, the infalling gas pulls the magnetic field
towards the center, strongly pinching the field lines near the
protostar. The parameters of the GS93 model are the spatial scale of
the core prior to collapse, $r_0$, and the nondimensional time elapsed
since the onset of collapse, $\tau$. The scale $r_0$ is defined in
terms of the effective sound speed $a_{\rm eff}$ (including thermal,
turbulent and magnetic support) and the initial magnetic field $B_0$ by
the expression $r_0=2a_{\rm eff}^2/(G^{1/2}B_0)$, where $G$ is the
universal gravity constant. The physical conditions in the core that
gave birth to IRAS~4A are difficult to derive. In particular, no Zeeman
measurement of the magnetic field strength in the surroundings of
IRAS~4A is available. Thus, we adopted the
``fiducial'' values of GS93, $a_{\rm eff}= 0.35$~km~s$^{-1}$, and
$B_0=30$~$\mu$G. Then $r_0=2.1\times 10^4$~AU, and the nondimensional
time $\tau$ is given by $\tau=t/t_0$ with $t_0=r_0/a_{\rm
eff}=2.9\times 10^5$~yr. The results can be easily scaled to different
values of $r_0$.

The only parameter of the S06 model is the Ohm's radius, a spatial
scale associated with Ohmic dissipation and defined as $r_{\rm
Ohm}=\eta^2/(2GM_\star)$, where $\eta$ is the Ohmic resistivity,
assumed to be spatially constant. With a non-zero resistivity, the magnetic
field lines strongly bent inward by the pull of the infalling gas relax
to an almost straight and uniform field configuration in a region of a
size $\sim r_{\rm Ohm}$, with the enclosed magnetic flux reduced
with respect to the field-freezing value (by a factor of $\sim 100$ at
$r=r_{\rm Ohm}$).  The value of $\eta$ depends on the ionization,
temperature, and chemical composition of the gas close to the source and
is not well-constrained.  Moreover, the value of $\eta$ is probably 
associated to macroscopic plasma instabilities rather than microscopic
collisional processes (Shu et al.~2007).  Estimates of $r_{\rm Ohm}$
are in the range 1--100~AU for a solar-mass star (S06).

\section{The method}
\label{method}

In spite of the idealization of the models, a comparison with the data
is not straightforward. First, the orientation in space of the models
is specified by two viewing angles, the position angle $\phi$ of the
polar axis with respect to a reference direction in the plane of the
sky ({\bf with} $\phi=0$ corresponding to a polar axis aligned
north-south), and the inclination $\psi$ of the meridional plane of the
model with respect to the plane of the sky (with $\psi=0$ corresponding
to a polar axis in the plane of the sky).  These orientation parameters
are well-constrained by the observations. In addition, the GS93 model
depends on two parameters, a spatial scale $r_0$ and time, whereas the
steady-state S06 model depends on a single parameter, the Ohm's radius
$r_{\rm Ohm}$. The latter in particular is not well-constrained by the
theory (S06).  Thus, the data must be compared with an extensive set of
model density and magnetic field distributions obtained for different
values of the model parameters and the orientation angles $\phi$ and
$\psi$.

First from a grid of density and magnetic field models we obtained a
large number of maps of Stokes parameters $I$, $Q$, and $U$, using the
same method as described in Gon\c calves et al.~(2005).  Once the
models were generated, they were convolved with the SMA interferometric
response.  This was done by converting the modeled map to visibilities
using the same distribution of visibilities in the $u,v$ plane and the
same $u$, $v$ weighting as the SMA observations of IRAS~4A. This
procedure was followed independently for the Stokes $I$, $Q$, and $U$.
Once the synthetic SMA--like Stokes $I$, $Q$, $U$ maps were obtained,
the polarization intensity and position angle were obtained in the same
way for the SMA data and for the models. Since this convolution is
nonlinear, it is necessary to scale the model intensity with that of
the observations, which is equivalent to selecting a specific value for
each map for the product $\kappa B(T)$, or, in other words, a mass
calibration factor.  This is performed with an iterative procedure.

The third and last step was to compare the modeled and observed
polarization angle at each observed position. A quantitive evaluation
of the goodness of the fit is given by the standard deviation of the
distribution of the residuals between the modeled and the measured
polarization angles. The continuum emission and the polarized intensity
predicted by the model were also used as additional constraints.

\section{Results}
\label{results}

\subsection{The GS93 model}
\label{gs93}

\begin{figure}[t]
\resizebox{\hsize}{!}{\includegraphics{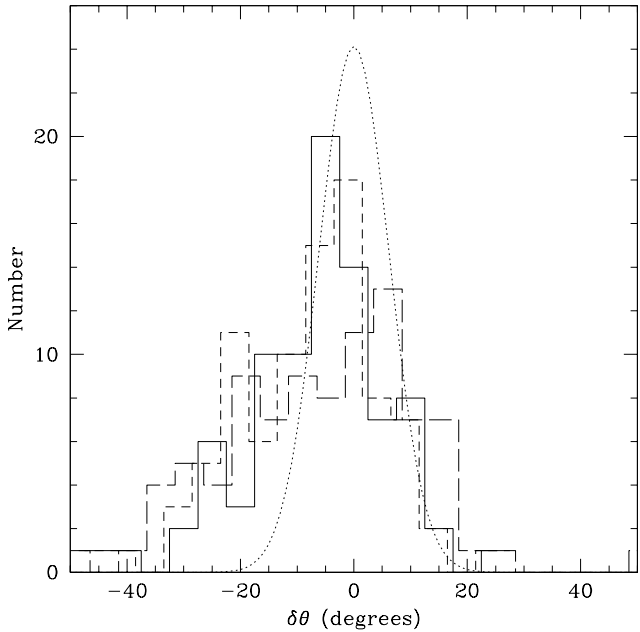}}
\caption{Histogram of residuals in polarization angles for the GS93
models.  {\em Solid} line, $\tau=0.3$; {\em short-dashed} line,
$\tau=0.5$~AU; {\em long-dashed} line, $\tau=0.7$ (histograms slightly
displaced for clarity). The {\em dotted} curve shows
the Gaussian distribution of the measurement uncertainty with 
$\sigma_{\rm instr.}=6.2^\circ$ (GRM06).}
\label{histo_gs93} 
\end{figure}

We have computed a total of 270 maps for $\tau=0.3$, 0.5, and 0.7, with
different values of $\phi$ and $\psi$. The best-fit models have
$\phi=50^\circ$--$60^\circ$, $\psi=0^\circ$--$30^\circ$, $r_0=
3.2\times 10^{17}$~cm, and $\tau=0.3$. Although the fit is not very
sensitive to the value of $\psi$, the best results are obtained for
small inclinations of the model with respect to the plane of the sky,
with the pseudo-disk seen almost edge-on.  In Fig.~\ref{histo_gs93} we
show the histograms of the residuals in polarization angles
(differences between predicted and observed values) for the three
values of $\tau$. The standard deviations are $\sigma=14.8^\circ$,
$15.6^\circ$, and $18.5^\circ$ for $\tau=0.3$, $0.5$, and $0.7$ (the
instrumental uncertainty on the polarization angles is $\sigma_{\rm
instr.}=6.2^\circ$, see GRM06).

\begin{figure}[t]
\resizebox{\hsize}{!}{\includegraphics[angle=-90]{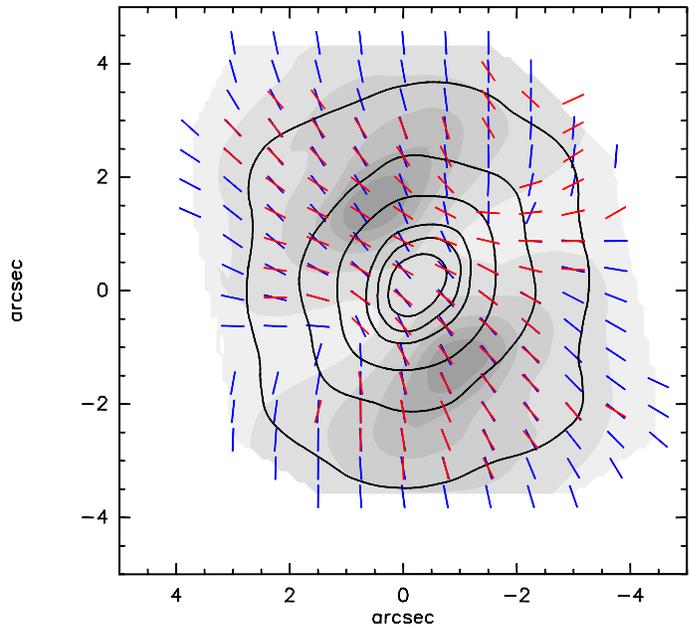}}
\caption{Synthetic maps for the GS93a model: continuum emission
(contours), polarized emission (greyscale), polarization degree
(segments). Red segments show the observed polarization vectors. Model
GS93 with $\tau=0.3$, $\psi=30^\circ$, $\phi=55^\circ$.}
\label{gs93_03} 
\end{figure}

Although the polarization angles and the intensity profile of IRAS~4A
can also be satisfactorily reproduced by models with $\tau=0.5$--0.7,
the isocontours of the emitted flux are flatter in these cases than
observed, and the polarized flux intensity not reproduced as well as
in the $\tau=0.3$ case. For $\tau=0.7$, neither the shape of the
isodensity contours nor the polarized intensity distribution can be
reasonably reproduced for any possible viewing angle. In
Fig.~\ref{gs93_03} we show a comparison of the modeled and observed
intensity and polarization map for the best-fit value $\tau=0.3$. For
this case, the largest contribution to $\sigma$ comes from a systematic
deviation of the polarization vectors on the top-left side of the map.

With our choice of parameters, the best-fit model with $\tau=0.3$
implies an age for this protostar of $t_\star=8.6\times 10^4$~yr and a
mass of $M_\star\approx a_{\rm eff}^3t_\star/G=0.8$~$M_\odot$.  The
mass of the infalling envelope within a radius $r=10^3$~AU from the
source is $0.5$~$M_\odot$, in agreement with the mass distribution
derived by Belloche et al.~(2006) from interferometric and single-dish
continuum observations ($0.7$ and $0.9$~$M_\odot$, respectively). From
the best-fit model, the magnetic flux enclosed in a radius $r=10^3$~AU
is $\Phi=2.4\times 10^{30}$~G~cm$^2$, so the mass-to-flux ratio of the
envelope plus star is $\sim 1.7$ times the critical value
$(M/\Phi)_{\rm cr}=(2\pi G^{1/2})^{-1}$, in agreement with the
independent estimate of GRM06. Since the effects of ambipolar diffusion
are negligible on the collapse time scale in the GS93 model, this value
characterizes the mass-to-flux ratio of the original 1.3 $M_\odot$ of
cloud core that have collapsed to form the protostar and the envelope.


\subsection{The S06 model}
\label{s06}

The S06 model is only applicable to a small region with respect to
$a_{\rm eff} t$ ($\approx 6000$~AU with the parameters of
Sect.~\ref{gs93}), and assumes steady-state over this region.  Thus, no
ages or stellar masses can be derived for this model.  In principle, a
comparison of the polarization data for IRAS~4A with the S06 model is
able to constrain the value of $r_{\rm Ohm}$ without relying on
uncertain estimates for the Ohmic resistivity.

\begin{figure}[t]
\resizebox{\hsize}{!}{\includegraphics{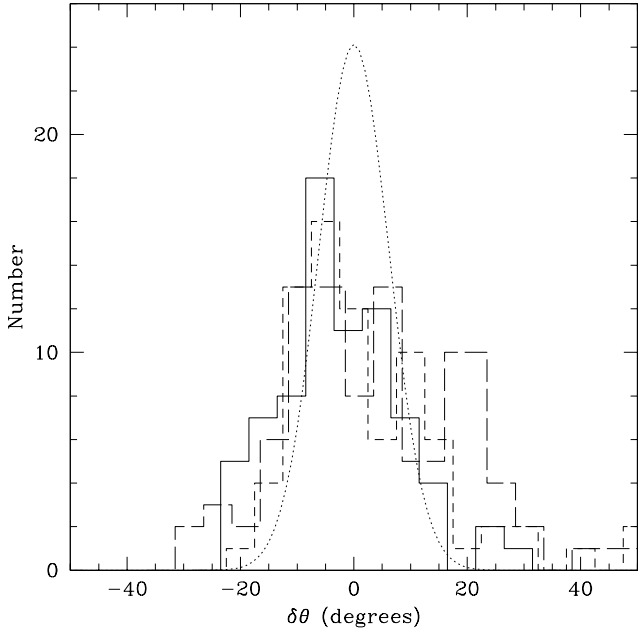}}
\caption{Histogram of residuals in polarization angles for the S06
models.  {\em Solid} line, $r_{\rm Ohm}=5$~AU; {\em short-dashed} line,
$r_{\rm Ohm}=50$~AU; {\em long-dashed} line, $r_{\rm Ohm}=500$~AU
(histograms slightly displaced for clarity). The {\em dotted} curve
shows the measurement uncertainty as in Fig.~\ref{histo_gs93}.}
\label{histo_s06} 
\end{figure}

\begin{figure}[t]
\resizebox{\hsize}{!}{\includegraphics[angle=-90]{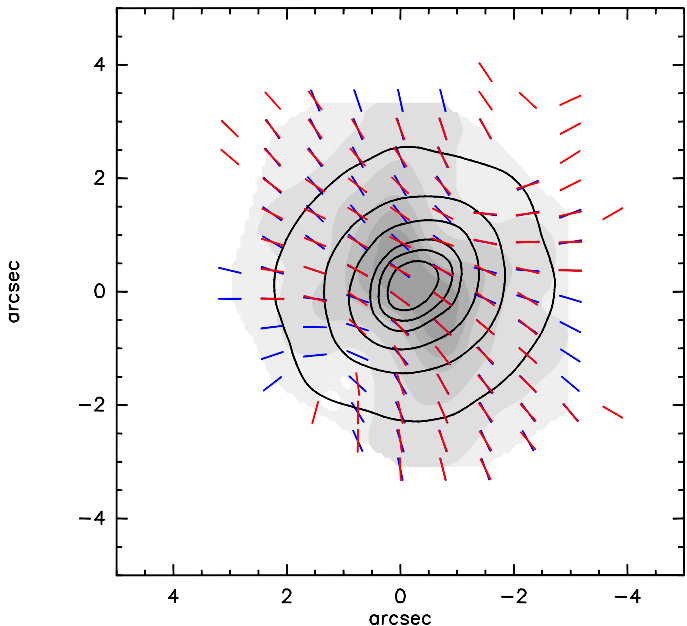}}
\caption{Synthetic maps for the S06 model: continuum emission
(contours), polarized intensity (contours and greyscale), polarization
degree (segments). Red segments show the observed polarization
vectors. Model with $r_{\rm Ohm}=5$~AU, $\psi=54^\circ$, $\phi=65^\circ$.}
\label{s06_5} 
\end{figure}

For the S06 model, we computed a total of $\sim 450$ synthetic maps
with different values of $\phi$ and $\psi$ and values of $r_{\rm Ohm}$
ranging from 5 to 500~AU.  Best-fitting models are obtained with
$\phi=60^\circ$--$65^\circ$ and $\psi= 25^\circ$--$55^\circ$. The
results for the GS93 model discussed in Sect.~\ref{gs93} are not very
sensitive to the value of $\psi$, and the observed polarization pattern
alone is not sufficient for constraining the model parameters. Models
with high values of $r_{\rm Ohm}$ ($r_{\rm Ohm}\approx 200$--500~AU)
are not able to reproduce the observed polarization pattern very well
or even the two-lobe structure of the polarized continuum emission.
The best fit to the polarization angles is obtained for models with
$r_{\rm Ohm} \la 50$~AU, suggesting that the actual value of $r_{\rm
Ohm}$ may be close to or below the spatial resolution of the
observations ($\sim 10^2$~AU). In fact, the two-lobe shape of the
polarized intensity emission evident in Fig.~\ref{iras4a} can only be
reproduced with very low values of the Ohm's radius, $r_{\rm
Ohm}=5$--50~AU, suggesting, in agreement with the findings of
Sect.~\ref{gs93}, that the magnetic field down to the scale of
resolution is dominated by a pinched, rather than uniform, component.
Figure~\ref{histo_s06} shows the histograms of the polarization angle
residuals for the models with $r_{\rm Ohm}=5$, 50 and 500~AU,
characterized by standard deviations $\sigma=11.7^\circ$, $13.8^\circ$,
and $22.6^\circ$. The polarization and intensity map for the case
$r_{\rm Ohm}=5$~AU is shown in Fig.~\ref{s06_5}.

\section{Discussion and conclusions}
\label{discussion}

From the discussion of Sect.~\ref{results}, it is clear that the
agreement of magnetic collapse models with the continuum polarization
data for IRAS~4A is very good, supporting the standard theoretical
scenario for the formation of low-mass stars from magnetized molecular
cloud cores.  The residuals distributions indicate that the best
agreement with the data is obtained with GS93 models with early ages
($\sigma=14.8^\circ$ for $\tau=0.3$) or for S06 models with low values
of the Ohm's radius ($\sigma=11.7^\circ$ for $r_{\rm Ohm}=5$~AU). The
low value of $r_{\rm Ohm}$ obtained from the best-fitting S06 model
($r_{\rm Ohm}=5$--50~AU) might also indicate that the steady-state
conditions assumed in that model have not yet been reached, and the
size of the region of strong magnetic dissipation is below the
resolution limit of the observation. At any rate, both models suggest
that the magnetic field lines in a region of radius $\sim 500$~AU from
the central source(s) are almost radial, as expected for collapse under
ideal MHD conditions.

The almost radial geometry of magnetic field lines in IRAS~4A is also
supported by the distribution of polarized intensity shown in
Fig.~\ref{iras4a}. In fact, only models with strong central
concentration of magnetic field lines reproduce the observed two-lobe
structure of the observed polarized intensity, with two large
``polarization holes'' in the midplane of the pseudo-disk, on both
sides of the central protostar.  Rather than a variable efficiency
alignement of dust grains, the presence and extent of these ``holes''
seem to indicate a combination of geometrical effect (many field lines
point to the observer in a strongly pinched geometry) combined with
beam dilution effects (the pinched region is smaller than the beam),
resulting in a cancellation of Stokes parameters along the
line-of-sight.

Some amount of field dissipation must have already occurred in the
central regions of IRAS~4A. The presence of two continuum peaks in
IRAS~4A separated by $1.8\arcsec$ (Looney et al~2000; Reipurth et
al~2002),  corresponding to 540~AU at the assumed distance of 300~pc,
suggests that some fragmentation of the pseudo-disk has already
occurred, leading to the formation of a (possibly bound) protobinary
system.  Without field dissipation, the catastrophic magnetic braking
associated to the concentration of magnetic fields in ideal-MHD
collapse calculation provides a fierce opposition to fragmentation and
to the formation of circumstellar disks (Galli et al.~2006; Hennebelle
\& Fromang~2008; Mellon \& Li~2008).

From the distribution of residuals of the fit, it is possible to
estimate the relative magnitude of the turbulent vs. the ordered
component of the magnetic field using the formula $|\delta{\bf
B}|/|{\bf B}|\approx (\sigma^2-\sigma_{\rm instr.}^2)^{1/2}$ (GRM06).
With the $\sigma$ obtained for the best-fitting GS93 and S06 models, we
derive $|\delta{\bf B}|/|{\bf B}|\approx 20\%$. This is probably an
upper limit on the intensity of the turbulent field. Thus, The magnetic
field morphology in IRAS~4A is consistent with the standard theoretical
scenario for the formation of low-mass stars from cores threaded by
ordered rather than turbulent magnetic fields.

\begin{acknowledgements}
We thank an anonymous referee for a careful review of the paper.
DG acknowledges support from the Marie-Curie Research Training Network
``Constellation'' (MRTN-CT-2006-035890); JMG from grants
AYA2005-08523-C03 (Ministerio de Ciencia e Innovaci\`on and FEDER) and
2005SGR00489 (Generalitat of Catalunya).
\end{acknowledgements}

\end{document}